\documentclass[10pt,sigplan,letterpaper,twocolumn]{article}
\usepackage[10pt,nocopyright]{sigmin}

\usepackage{algorithmic}
\usepackage{textcomp}
\usepackage{xcolor}
\usepackage{filecontents}
\usepackage{listings}
\usepackage{xcolor}
\usepackage{colortbl}
\usepackage{balance}
\usepackage{graphicx}
\usepackage[hyphens]{url}
\usepackage[hyphenbreaks]{breakurl}
\usepackage{hyperref}
\usepackage{threeparttable}
\usepackage[linesnumbered,boxed,lined,ruled,vlined]{algorithm2e}
\usepackage{setspace}
\usepackage[numbers]{natbib}
\usepackage{diagbox}

\usepackage{subfig}
\usepackage{wrapfig}
\usepackage[export]{adjustbox}
\usepackage{times}
\usepackage[scaled=0.95]{inconsolata}
\usepackage{ifsym}
\usepackage{breqn}
\usepackage{bbding}
\usepackage{booktabs}
\usepackage{multirow}
\usepackage{amssymb}
\usepackage{enumitem} 

\usepackage{mathptmx}
\usepackage{mathtools}

\usepackage{mdframed}

\usepackage{multicol}
\usepackage{rotating}
\usepackage{listings}
\usepackage[compact]{titlesec}
\titleformat*{\section}{\large\bfseries}
\titleformat*{\subsection}{\normalsize\bfseries}
\titleformat*{\subsubsection}{\normalsize\bfseries}
\titlespacing*{\section}{0pt}{*2}{*1}
\titlespacing*{\subsection}{0pt}{*1.5}{*0.8}

\usepackage{authblk}
\usepackage{longtable}
\usepackage{supertabular}
\usepackage{pifont}
\usepackage{threeparttable}
\usepackage{makecell}
\usepackage{soul}

\usepackage{tikz}

\let\origunderscore\_
\DeclareRobustCommand{\code}[1]{{\texttt{\renewcommand{\_}{\origunderscore\hspace{0pt}}#1}}}

\newcommand{\sys}{Tide}
\newcommand{\proxy}{Anchor}
\newcommand{\rssint}{backing memory}

\newcommand*\circled[1]{\tikz[baseline=(char.base)]{
            \node[shape=circle,draw,inner sep=0.1pt, minimum size=9pt] (char) {\footnotesize #1};}}

\newcommand{\stitle}[1]{\vspace{0.5ex}\noindent{\bf #1}}

\newif\ifshowLLMsuggestions
\showLLMsuggestionstrue

  {\begin{list}{\labelitemi}{\itemsep1pt \topsep2pt \parsep0.00in
  \partopsep=0pt \leftmargin1.4em}}%
  {\end{list}}

\AtBeginDocument{%
  \providecommand\BibTeX{{%
    \normalfont B\kern-0.5em{\scshape i\kern-0.25em b}\kern-0.8em\TeX}}}

\begin{document}


\title{\Large \sys{}: Reclaiming Phased Memory in Agent MicroVMs}

\author{
Yiyang Wu,
Chengfan Liao,
Jinyu Gu
\\
\emph{Institute of Parallel and Distributed Systems, Shanghai Jiao Tong University}\\
}


\pagestyle{plain}

\date{}
\maketitle

\frenchspacing


\begin{abstract}
Cloud agents run each task in an isolated MicroVM. The trouble is the harness
loop inside that guest: the harness is nearly idle while it waits on the
model, then usage rises on a tool whose size is known only at run time,
which complicates the memory management from the host.
Existing approaches infer reclaim targets from access frequency
and memory footprint while the guest allocator places the
idle harness and the short-lived tool on the same pages.
Therefore, reclamation either selects the wrong pages, fixes on an incorrect capacity or
partially reclaims a huge page, splitting its transparent huge page (THP).

This paper proposes \sys{}, a proactive memory reclamation technique for
agent MicroVMs.
Our key insight is that the harness already knows each phase's memory
intent---which memory, when, and whether its contents must survive---so the
guest can report that intent as an exclusive guest-physical region for
the host to reclaim directly.
To realize this insight, we introduce (1)~an abstraction called arena and a set of
user-space API that can manage memory with an exclusive intent.
(2)~a guest-kernel allocator that places the
realize the allocation that groups exclusive, huge-page-aligned extents; and (3)~a
hypervisor extension that resolves those extents to host backing and applies
the reported action, leaving guest capacity unchanged.
Experiments on recorded agent trajectories show that \sys{} outperforms
state-of-art mechanism while incuring low overhead.
\end{abstract}

\section{Introduction}
\label{sec:intro}

Agents are moving into the cloud. Applications such as Claude
Code~\cite{anthropicclaudecode}, OpenCode~\cite{opencode},
Codex~\cite{openaicodex}, and OpenClaw~\cite{openclaw} translate
natural-language goals into sequences of model interactions and tool
executions, and platforms now run each task inside an isolated
MicroVM~\cite{agache2020firecracker,katacontainers}. The guest hosts the
\emph{harness}---the runtime that assembles context, calls the model, and
executes tools---together with the processes those tools spawn, while model
inference stays in a separate service. A host serves many tenants, so each
agent gets its own guest kernel, and the cost of that isolation is physical
memory held for the life of the task.

Four properties of memory properties matter for hosting agentic MicroVMs. 
First, demand follows the phase of the loop. The guest is nearly idle while the harness waits on the
model, and usage rises when a tool runs locally. Second, the model chooses the
next tool only at run time, so the size of each resource burst is hard to predict.
Third, harness pages and tool pages share one guest and follow different
lifetimes. Harness state stays live for the whole run, while a tool's
private pages last only until its process tree exits. Lastly, host
backing follows the peaks and stays after guest usage has fallen,
including memory that can already be reclaimed.

However, existing memory management techniques still miss these properties
on different dimensions. Host swapping ranks pages by access frequency,
so it cannot separate idle-but-live harness pages from already-freed
tool pages and may write those freed pages out in unnecessary I/O.
Virtio-balloon~\cite{virtio12} and
virtio-mem~\cite{hildenbrand2021virtiomem} reclaim by shrinking guest
capacity to a target, but a target sized for the wait may not fit the
next burst causing severe direct guest reclaim.
On top of that, the guest allocator places independent lifetimes in the
same host huge pages, which causes reclaiming one lifetime
splits huge-page mappings the other still uses.

The root cause is that the harness already knows each phase's
\emph{memory intent} from its control flow but the harness has no
capability to report it. An intent names a scope, a
trigger, and a preservation rule. During a
long model wait, the scope is the harness and its persistent services,
the trigger is the wait, and the rule is to page out state that is idle
but live. When a tool's process tree exits, the scope is that
invocation, the trigger is the exit, and the rule is to discard its
private state sooner, however much memory the tool used.
Two gaps keep that intent from driving reclamation. 
On the guest side, every phase allocates from the same buddy allocator, 
so pages of different intents are not distinguishable as regions and share host huge pages.
On the host side, memory intent can only be inferred from access frequency
and memory footprint and the guest has no way to report the intent across
the guest--host boundary.

We present \sys{}, which closes both gaps with one abstraction, the
\emph{arena}. Its user-space API places each memory intent in its own
guest-physical region, and operating on one arena leaves the others
untouched. Harness state, including persistent services, lives in a
dedicated arena~H, and each tool invocation lives in its own arena~T.
Huge-page-aligned extents from a shared CMA pool give each arena
exclusive host huge pages. User-space operations then can tell the host
how to reclaim that region's backing without shrinking guest capacity.
Since the harness itself cannot issue arena operations during a model wait or
tool invocation, we design \proxy{}, a user-space component that stays outside
H and T, delegates model requests and tool execution, and issues arena
operations at each phase boundary.
The guest reports the operation and the arena's guest-physical ranges
to the hypervisor over a virtio channel. The hypervisor resolves those
ranges to host backing through its own translation tables and applies the
corresponding memory action while the host kernel stays unmodified.

We implement \sys{} in Linux, Firecracker, and OpenCode, OpenClaude, and
Codex, with \proxy{} written in Rust. On the longer SkillsBench tasks,
\sys{} stays within 1.7\% of the baseline.
On six Terminal-Bench tasks~\cite{merrill2026terminalbench} running on
OpenCode, OpenClaude, and Codex, \sys{} reduces total \rssint{} by 45.3\%,
41.0\%, and 34.2\% relative to baseline and by 46.8\%, 39.3\%, and 32.9\%
relative to HyperAlloc, with harness-time increases of 5.1\%, 1.1\%, and
1.9\%.

\section{Background and Motivation}
\label{sec:motivation}

\subsection{Cloud Agents in MicroVMs}
\label{sec:cloud-agents}
Agents are moving to the cloud. Model and cloud providers now offer hosted
agents and agent platforms, including OpenAI Codex~\cite{openaicodex}, Claude
Code~\cite{anthropicclaudecode}, Amazon Bedrock AgentCore~\cite{awsagentcore},
Google Vertex AI Agent Engine~\cite{googleagentengine}, Microsoft Foundry
Agent Service~\cite{microsoftfoundryagents}, and Tencent WorkBuddy Managed
Agents~\cite{tencentworkbuddy}. Apart from interactive usage or automatic cloud tasks,
the appeal of hosting is that it decouples the agent from the user's device. 
A user can launch many tasks at once and
let each run for hours, which is impractical on a laptop that must stay
online; hosted agents keep working in the background and deliver results
when they finish. The platform, in turn, provisions execution environments
consistently at scale and keeps whatever the agent executes inside a sandbox
instead of on the user's machine.

On these platforms, each agent runs in its own MicroVM. A host serves agents
from many tenants, so the boundary between them must be stronger than a
shared-kernel container can offer. MicroVMs such as
Firecracker~\cite{agache2020firecracker} and Kata
Containers~\cite{katacontainers} give each agent a dedicated guest
kernel, and they are now the sandbox of choice on agent platforms such as
AgentCore~\cite{awsagentcoreruntime} and E2B~\cite{e2b}. The MicroVM hosts
the \emph{harness}, a dedicated runtime that assembles context, calls the LLM service,
and executes the tool calls the model requests, together with the workspace
it operates on and the tool processes it spawns. Model inference itself runs outside the VM in a separate
LLM service.

\subsection{Why Reclamation Falls Short in Agent MicroVMs}
\label{sec:reclamation-limitations}
A host today can take memory back from a VM in two ways, swapping out cold
pages, or shrinking the guest's capacity. 
On an agent VM, swapping and shrinking do reduce the host's footprint
but they fail to notice the memory's lifetime and demand burst inside the harness
loop, resulting in performance degradation or even service loss. On top of that,
even if the host knew which memory to reclaim, the placement of that memory
could still degrade performance. 

\stitle{Swapping Lacks Lifetime Awareness.}
Host swapping reclaims pages that look cold: Linux LRU~\cite{gorman2004linuxvm}
and Multi-Generation LRU~\cite{linuxmglru} rank pages by access history, and
programmable eviction policies~\cite{zussman2025cache} still work from access
signals alone. In an
agent VM, however, cold pages come in two kinds that access history cannot
tell apart. Idle-but-live pages belong to a harness waiting on the model or
to a service between queries; their contents will be needed again. Dead pages
belonged to a tool whose process tree has exited; their contents will never
be needed. Swapping treats both the same way and is wrong for both: preserving dead
contents spends swap I/O and capacity on data nobody will read, while paging
out live contents pays off only if the idle interval is long enough to
amortize the swap-in on the next access. Access recency reveals neither when
a tool's lifetime ended nor how long an idle interval will last.

\stitle{Adjustment Lacks Demand Awareness.}
Two mechanisms, virtio-balloon~\cite{virtio12} and
virtio-mem~\cite{hildenbrand2021virtiomem}, shrink a VM's capacity so that
the host can reuse the difference, which
requires choosing a target. On an agent VM there
is no safe static target: a capacity that fits the waiting harness does not
fit the next tool, and a tool that allocates beyond it drives the guest into
sustained direct reclaim, stalling or failing. Faster or cheaper
adjustment~\cite{wrenger2025hyperalloc} does not help: however quickly the
balloon inflates, the tool still runs in a VM sized for the wait. A safe
target would have to be raised before each burst but not until the tool
is actually invoked can the controller observe the memory fluctuation.

\stitle{Reclamation Lacks Placement Awareness.}
Suppose the host knew exactly which pages were dead and left ample capacity,
even with that knowledge, ciscarding them could still slow down live state
as a side effect, because of where they sit.
Virtualized address translation relies on host Transparent Huge
Pages~\cite{linuxthp} (THPs, typically 2\,MB) to keep two-dimensional page
walks~\cite{bhargava2008twodimensional} cheap, yet the guest
allocator places pages without regard to lifetime or to host huge-page
boundaries. A 2\,MB host page therefore routinely mixes long-lived harness
and kernel pages with short-lived tool pages, and one tool's pages scatter
across many such huge pages. The problem is twofold. On one hand,
HyperAlloc~\cite{wrenger2025hyperalloc} and free-page reporting~\cite{virtio12},
newer mechanisms that proactively report freed guest memory can be blocked
because scattering leaves those pages only partly free.
On the other hand, reclaiming the dead pages of a mixed huge page
forces a bad choice. The host either splits the mapping, leaving the live
pages on 4\,KB mappings until \texttt{khugepaged} re-collapses them, or it
keeps the dead backing resident so the huge-page mapping stays intact. 
No data is lost either way,
but reclaiming correctly identified dead memory now has a side effect.
It either slows live state or leaves dead backing on the host. Reclaiming
lifetimes independently requires placing them apart at
huge-page granularity, which only the guest allocator can do.


In each case, the missing piece is something the harness already has: when
it waits, when a tool's process tree exits, and which pages were that tool's.
The host also needs a placement that lets it act on this knowledge without
side effects on the live state around it. This paper bridges this gap.


\section{Key Ideas: Memory Intents and Arenas}
\begin{figure*}[t]
    \centering
    \captionsetup{font=normalsize}
    \captionsetup[subfloat]{font=normalsize}
    \subfloat[System architecture.\label{fig:overview-architecture}]{%
        \begin{minipage}[b]{0.48\textwidth}
            \centering
            \includegraphics[width=\linewidth]{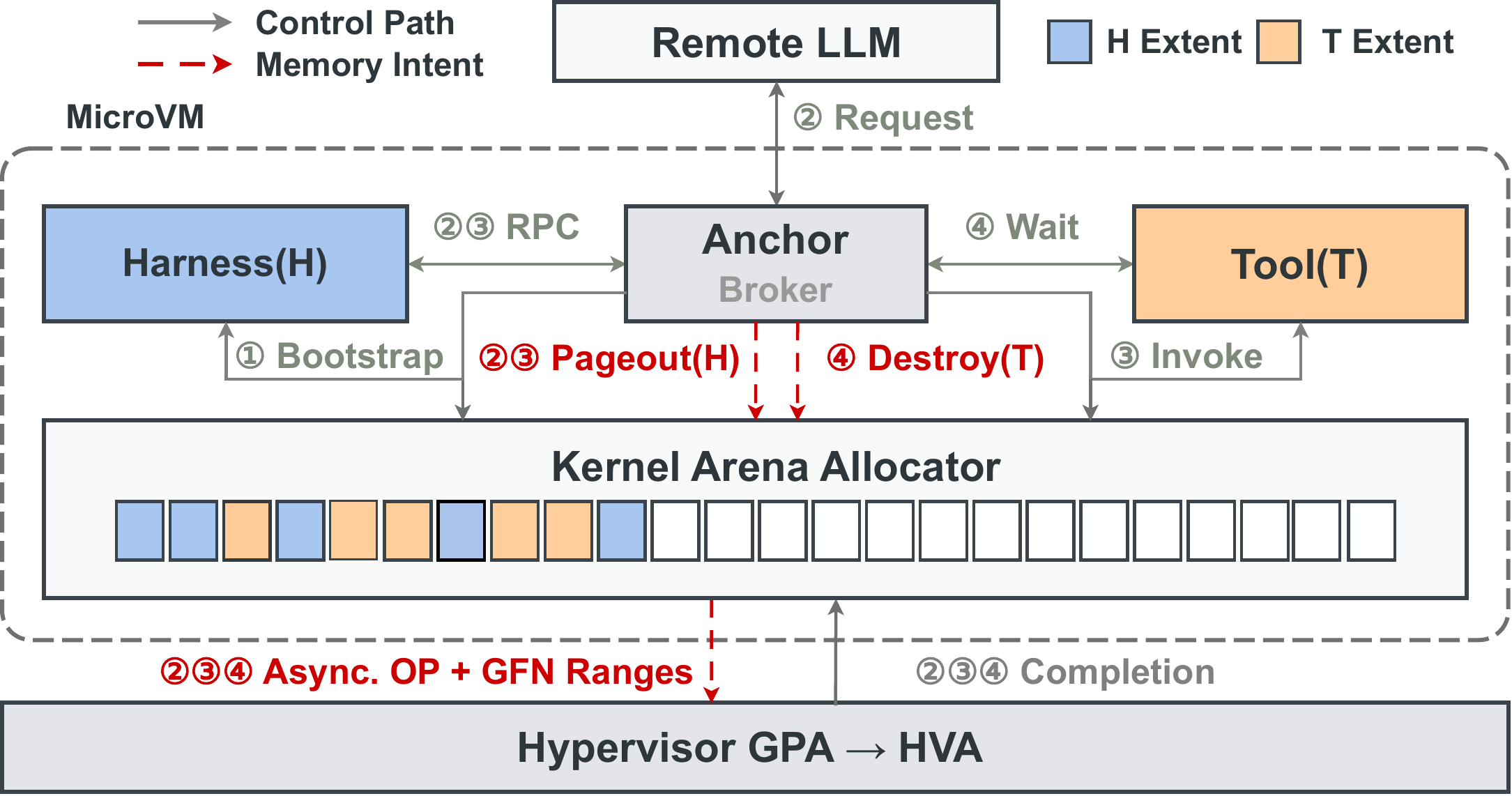}
        \end{minipage}%
    }
    \hfill
    \subfloat[Harness loop.\label{fig:overview-interface}]{%
        \begin{minipage}[b]{0.48\textwidth}
            \centering
            \includegraphics[width=\linewidth]{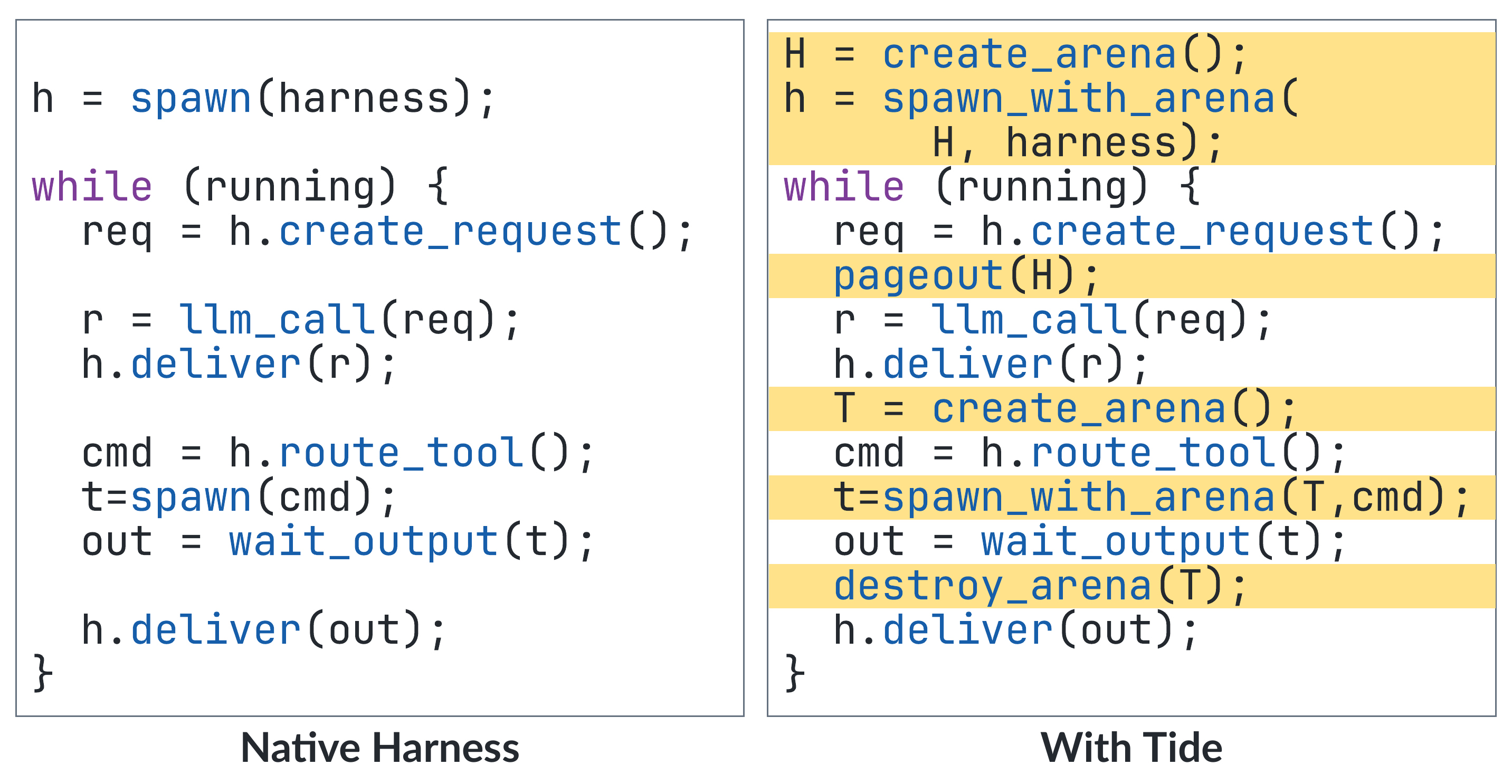}
        \end{minipage}%
    }
    \caption{\sys{} architecture and the harness loop. (a)~\proxy{} in guest
    user space turns phase and lifetime boundaries into memory intents; the
    guest kernel resolves each intent into the extents of one arena; and the
    hypervisor acts on their host backing. (b)~The loop without and with
    \sys{}: \code{create_arena} establishes H and T;
    \code{spawn_with_arena} starts each process in its arena;
    \code{pageout(H)} on each model wait;
    \code{pagein(H)} swaps that backing back in before the harness resumes,
    so the next access does not fault; and \code{destroy_arena(T)} after
    the tool exits.}
    \label{fig:overview}
\end{figure*}

\label{sec:overview}

The three failures in Section~\ref{sec:reclamation-limitations} share a root
cause: the host has to \emph{infer} what guest memory means, and the guest
does not know how to place the memory in the reclamation-friendly way. 
Swapping guesses lifetimes from access history and
capacity adjustment guesses demand from a footprint, and both may guess wrong
because they observe only the traces those events leave behind rather than the events themselves.
The agent harness knowes it from the events directly: it knows when it starts
waiting on the model, when it launches a tool, and when that tool's process
tree exits.
Our system, \sys{}, therefore reverses
the direction of information. Instead of the host inferring from below, the
guest reports \emph{memory intent} from above---which memory can be
reclaimed, when, and whether its contents must survive---and the host acts on
it under its own policy.

The agent loop exposes exactly two such opportunities, one for each memory lifetime. 
When the harness begins a model wait, its state is idle but live,
and the intent is \code{pageout} meaning that
the harness's backing can be swapped out, because they
will only be needed when the response arrives. When a tool's process tree exits,
its private memory is dead, and the intent is \code{destroy} meaning 
that tool's backing can be discarded soon. Each intent thus names a scope (the
harness, or one tool invocation), a trigger (the wait begins, the tree
exits), and a preservation rule. Neither shrinks the guest's capacity: a tool
always allocates freely, and reclamation acts only on a waiting harness or an
exited tool, so the target problem of
Section~\ref{sec:reclamation-limitations} never arises.

Carrying these intents from user mode to the host meets three gaps.
First, nothing in the guest names the memory one invocation owns, as its pages are
interleaved with every other allocation.
Second, the host reclaims guest-physical ranges, while a process tree names only
virtual addresses, so collecting those ranges means a page-table walk over
thousands of scattered pages.
Third, those pages share host huge pages with other allocation domains, so
reclaiming them splits mappings that live state still uses.

\sys{} closes the gaps with an \emph{arena}: an exclusive set of
\emph{extents}, contiguous 2\,MB-aligned guest-physical ranges drawn on demand
from a shared pool and bound to one lifetime.
In a VM, untouched extents cost the host nothing, because it backs only the
pages the guest touches~\cite{zeng2026compactionfree}, so each lifetime can
keep its own ranges. Aside from the zero-cost, the gap is thus closed with this simple abstraction.
First, one arena per lifetime names the memory that lifetime owns.
The harness occupies a persistent arena~H, and \code{spawn_with_arena} starts
each tool in its own arena~T, so the tree allocates only from T and
destroying T leaves H untouched.
Second, the arena allocator records those guest-physical ranges as it
allocates them, so the report to the host is that extent list and needs no
page-table walk.
Third, each extent is exclusive and aligned to 2\,MB, so discarding T
returns whole host huge pages and leaves the huge pages that back H intact.

\section{Design}
\label{sec:design}

\subsection{Arena API}
\label{sec:arena-api}

Table~\ref{tab:arena-abi} lists the Arena API provided for the userspace applications.
Each function operates on an arena by its file descriptor.
The \code{spawn_with_arena} binds a new process to an arena. From then on,
supported allocations come from that arena and forked children inherit the
binding, while allocations and cached pages that already exist keep their
owners. The runtime therefore spawns each process into its arena before
that process initializes, so initialization allocates the memory that
\code{pageout} and \code{destroy_arena} can later reclaim.

\begin{table}[t]
    \centering
    \small
    \caption{Guest ABI operations. Parameters shown are the principal inputs.}
    \label{tab:arena-abi}
    \begin{tabular}{@{}p{0.45\columnwidth}p{0.45\columnwidth}@{}}
    \toprule
    Operation & Contract \\
    \midrule
    \code{create_arena()} & Create an arena. \\
    \code{spawn_with_arena(fd, elf)} & Start a process with an arena. \\
    \code{pageout(fd)} & Pageout the arena. \\
    \code{pagein(fd)} & Pagein the arena. \\
    \code{destroy_arena(fd)} & Close the arena after checks. \\
    \bottomrule
    \end{tabular}
\end{table}

The \code{pageout} requests a content-preserving swapout of a live arena's
host backing, typically during the harness's long waits, until
the next harness access swaps them back in on demand.
The \code{pagein} requests a proactive swap-in of a live arena's
host backing, typically once the next harness access is imminent,
so that access does not incur major faults.
The \code{destroy_arena} discards an arena's host backing after checks confirm
that nothing remains attached. It is still required after
process exit, since the latter only unmaps the address space and leaves the arena and
its backing in place. Once the hypervisor acknowledges the discard,
the guest kernel can safely reuse its memory.

\subsection{Architecture and Workflow}
\label{sec:overview-workflow}

\stitle{Architecture.}
Figure~\ref{fig:overview-architecture} organizes \sys{} into three layers.
Guest user space contains a long-running harness in arena~H, tool process
trees in separate arenas~T, and \proxy{}, a daemon that runs outside those
arenas. The guest kernel contains the arena allocator, which manages a shared pool
of exclusive physical extents, and a virtio interface that reports each
memory intent together with its affected ranges.
The hypervisor contains the corresponding virtio backend, which resolves
those guest-physical ranges to host backing, applies the requested
reclamation, and reports completion.

\stitle{Workflow.}
The workflow is then built around the observation in Section~\ref{sec:overview}.
Originally, such outward operations are performed by the harness itself, 
however as we proceed to swapping out the harness state, we need a proxy
to delegate the outward calls, and we call that proxy \proxy{}.
\proxy{} thus handles RPC from the harness to delegate model communication and
tool invocation and turns each execution phase into an arena operation that notifies the guest kernel.
Figure~\ref{fig:overview-architecture} shows the guest user-space
components that carry it with detailed
control and intent flow steps while Figure~\ref{fig:overview-interface} shows the
same stages on an agent control flow and how \sys{} inserts the arena operations between each step.

\circled{1}~\textbf{Bootstrap.}
To page out the harness, \proxy{} must hold its arena handle and know when
the harness is waiting and how to suspend it.  In addition to
ordinary startup, \sys{} therefore runs a bootstrap step that supplies
these. During bootstrap, \proxy{} creates a persistent arena~H and starts
the harness in it with \code{spawn_with_arena} 
while retaining its arena handle and its cgroup handle.
Once initialization finishes, the harness opens a persistent
connection to \proxy{}, and later model communication and tool invocation
pass through it, so \proxy{} sees each wait as it begins. \proxy{} itself
runs outside H, and its communication and result buffering stay out of the
memory that \code{pageout} will affect. Moreover, as the control flow of the harness
is now partly delegated, the harness cannot act on user signals such
as \code{SIGTERM} on itself. 
Therefore \proxy{} intercept them and decide whether to stop
the in-flight model or tool call or to stop the harness.

\circled{2}~\textbf{Model request.}
When the harness delegates a model request to \proxy{} through socket RPC,
\proxy{} forwards the request to the remote LLM and waits for the response.
However, \code{pageout(H)} is safe only when every agent that can touch H is waiting.
Parent and subagents share the same arena ~H, 
and one agent may have several model or tool calls in flight.
To resolve this, we design a suspension broker inside \proxy{} that can validate the condition. 
The broker records the live agents, and counts how many of each agent's calls have left
the harness. Only if the condition is met,
the broker can safely suspend the harness through its retained cgroup handle
and allowes \code{pageout(H)} so that H's host backing can be reclaimed into swap
to save host's memory. To avoid unnecessary swapping on short request,
\proxy{} waits for a grace period before issuing \code{pageout(H)}.
Each returning call releases one count and
the harness is resumed by the \proxy{} once any response is forwarded from \proxy{}.
As an optimization, \proxy{} requests \code{pagein(H)} when the LLM's
reasoning stage ends. The hypervisor tracks H's active extents and swaps
in only those the next harness access needs, so the resume does not incur
major faults.
 
\circled{3}~\textbf{Tool invocation.}
If the model requests a tool after model request,
the harness also needs to delegate that tool invocation to
\proxy{}. Compared with regular tool invocations,
\proxy{} creates a separate arena~T and starts the tool with
\code{spawn_with_arena} so that its descendants inherit the arena binding. This
separates the invocation's allocations from H instead of letting a tool
forked directly from the harness inherit H. While the tool runs, the harness
waits, allowing the runtime to issue \code{pageout(H)} under the same
grace-period and suspension conditions. For performance, \proxy{} applies
a heuristic split, where short-running, low-cost commands such as \code{ls},
\code{cd}, and \code{pwd} skip arena management and are invoked directly by
\proxy{} to avoid the cost of arena binding.

\circled{4}~\textbf{Tool completion.}
When the tool's process tree exits, this invocation's private memory is
dead, but exit only unmaps its address space, so arena~T and its host
backing remain until \proxy{} reclaims them. As The harness still needs the
tool's output after that reclamation, the tool writes stdout and stderr
to an anonymous pipe and \proxy{} buffers that output outside~T.
\proxy{}, which holds~T, then issues \code{destroy_arena(T)}.
Once cleanup checks confirm that nothing remains attached, the guest kernel
submits the discard of~T.
After the completion, the harness may steps~\circled{2}--\circled{4} until the model
returns a finish signal or the user intervenes with an key-interrupt signal.
\section{Evaluation}
\label{sec:eval}


\subsection{Methodology}
\label{sec:eval-methodology}

\stitle{Testbed.}
Our testbed uses an Intel Xeon Gold 5317 CPU at 3.00\,GHz, 256\,GB of RAM,
and a 512\,GB Samsung MZQL2960HCJR-00A07 NVMe SSD.

\subsection{Application Performance}
\label{sec:eval-application}



\begin{figure}[t]
    \centering
    \includegraphics[width=\columnwidth]{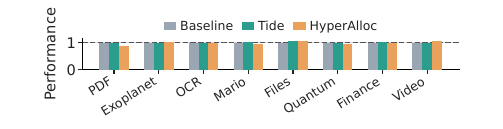}
    \caption{SkillsBench ground-truth execution relative to the baseline.
    Higher is better.}
    \label{fig:eval-skillsbench}
\end{figure}


\stitle{SkillsBench Ground-Truth Execution.}
Figure~\ref{fig:eval-skillsbench} compares eight offline \texttt{solve.sh}
tasks, counted when the command exits successfully and the expected outputs
are present. 
These are typical agent tasks, spanning document editing, file organization,
media processing, and numerical or tabular analysis.
On the six tasks whose baseline median exceeds 1\,s, \sys{}
stays within 1.7\% of the baseline. The largest change is Finance, from
15.09 to 14.84\,s, 1.7\% sooner. Relative to HyperAlloc, \sys{} is
1.2--5.3\% faster on OCR, Finance, Quantum, and Mario, and 3.3--5.1\% slower
on Exoplanet and Video. PDF and Files finish in under a second, and their
median gaps stay within about 50\,ms, the runner's completion-polling
granularity. PDF's baseline of 0.364\,s rises to 0.414\,s under HyperAlloc. 

\subsection{Agentic Memory Efficiency}
\label{sec:eval-memory}

\begin{figure}[t]
    \centering
    \includegraphics[width=\columnwidth]{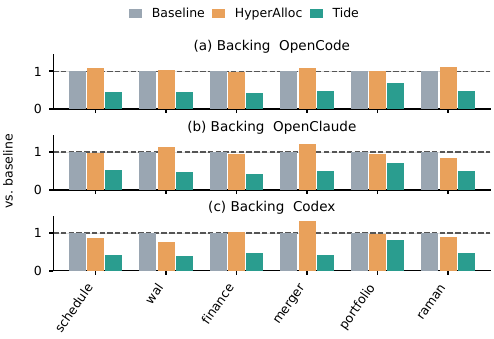}
    \caption{Terminal-Bench \rssint{} relative to the same-harness baseline;
    lower is better.
    (a)~OpenCode. (b)~OpenClaude. (c)~Codex.}
    \label{fig:eval-terminalbench-backing}
\end{figure}


Figure~\ref{fig:eval-terminalbench-backing} reports the \rssint{} integral
and harness time on six Terminal-Bench tasks~\cite{merrill2026terminalbench},
one trajectory each on OpenCode, OpenClaude, and Codex.
They range from calendar scheduling and log recovery to document extraction,
table merging, and numerical fitting: Schedule, WAL, Finance, Merger,
Portfolio, and Raman.
Backing memory is the harness's share of host memory after removing the
floor present at the start of replay.
The baseline and \sys{} run in Firecracker, with free-page reporting on the
baseline; HyperAlloc runs in QEMU.
Relative to baseline, total \rssint{} is 45.3\%, 41.0\%, and 34.2\% lower
and harness time is 5.1\%, 1.1\%, and 1.9\% longer.
Relative to HyperAlloc, the reductions are 46.8\%, 39.3\%, and 32.9\%, and
harness time is 2.7\% longer on OpenCode, 1.5\% shorter on OpenClaude, and
0.2\% shorter on Codex.
The added time is demand paging: synchronous \code{pageout} bypasses the
swap cache, so every fault goes through the I/O stack.
These tools consume little memory, so HyperAlloc stays near the baseline,
2.9\% higher, 2.8\% lower, and 1.9\% lower.
The duration-weighted average falls by 47.9\%, 41.7\%, and 35.4\%.


\section{Related Work}
\label{sec:related-work}

\stitle{Memory Elasticity in Virtualized Systems.}
A VM provisioned for its peak leaves memory idle whenever the footprint falls.
Vertical elasticity adjusts the allocation to measured demand~\cite{molto2016verticalelasticity}.
Squeezy returns gigabytes from a serverless VM in under a second~\cite{nikolos2026squeezy}.
HyperAlloc lets the hypervisor reclaim guest frames that are already free~\cite{wrenger2025hyperalloc}.
Puffer scales a function MicroVM by resizing its memory in blocks~\cite{fan2026puffer}.
Each of them reacts to a runtime signal, such as the footprint or the set of free frames.
\sys{} acts on an application-reported intent, which names the region and whether its contents must survive.

\stitle{Semantic-Aware Memory Management.}
Applications and the OS evolve separately, so the OS sees little of what a workload's memory means.
M3 coordinates application-specific reclamation across software layers~\cite{lion2021m3}.
TierVM delegates fine-grained tiering to the guest, using knowledge available only inside the VM~\cite{xing2025tiervm}.
Demeter and vtism likewise drive tiering from guest- or VM-level information~\cite{hu2025demeter,lu2025vtism}.
Disaggregated memory chooses a transfer size from observed spatial locality~\cite{koh2019disaggregated}.
These systems guide reclamation, placement, or movement from what they observe.
\sys{} reports the lifetime itself: which region, when, and whether its contents must survive.

\section{Conclusion}
\label{sec:conclusion}

We propose \sys{}, a proactive memory reclamation technique for agent
MicroVMs that reclaims idle harness backing during a model wait and dead
tool backing after the tool exits.
The key insight is that the harness already knows each phase's memory
intent, so the guest can report that intent as an exclusive guest-physical region, thereby
letting the host reclaim the region directly.
The design spans the harness, the guest kernel, and the hypervisor, and
end-to-end replay reduces backing memory substantially at low overhead.


\balance

\small{
\bibliographystyle{abbrv}
\bibliography{paper}
}

\end{document}